\documentclass[draft]{agujournal2019}
\usepackage{url} 
\usepackage[inline]{trackchanges} \usepackage{soul}
\usepackage{comment}

\draftfalse

\journalname{Space Weather}

\begin{document}

\title{Using Solar Orbiter as an upstream solar wind monitor for real time space weather predictions}

\authors{R. Laker\affil{1},
T. S. Horbury\affil{1},
H. O'Brien\affil{1},
E. J. Fauchon-Jones\affil{1},
V. Angelini\affil{1},
N. Fargette\affil{1},
T. Amerstorfer\affil{2},
M. Bauer\affil{2},
C. M\"ostl\affil{2},
E. E. Davies\affil{2},
J. A. Davies\affil{3},
R. Harrison\affil{3},
D. Barnes\affil{3},
M. Dumbovi\'c\affil{4}
\\}

\affiliation{1}{Imperial College London, Blackett Laboratory, South Kensington, SW7 2AZ}
\affiliation{2}{Austrian Space Weather Office, GeoSphere Austria, Reininghausstra{\ss}e 3, 8020 Graz, Austria}
\affiliation{3}{RAL Space, STFC Rutherford Appleton Laboratory, Harwell Campus, Didcot OX11 0QX, UK}
\affiliation{4}{Hvar Observatory, Faculty of Geodesy, University of Zagreb, Croatia}

\correspondingauthor{Ronan Laker}{ronan.laker15@imperial.ac.uk}

\begin{keypoints}
\item Real time data from Solar Orbiter and STEREO-A were used to predict the arrival time of two coronal mass ejections before arrival at Earth
\item For one event, in situ measurements at 0.5\,au were used to reduce the error in arrival time from 10.4\, to 2.5\,hours with the ELEvoHI model
\item The in situ Bz profile was comparable to the geomagnetic response at Earth, despite being separated by 0.5\,au and 10$^{\circ}$ longitude
\end{keypoints}

\begin{abstract}
Coronal mass ejections (CMEs) can create significant disruption to human activities and systems on Earth, much of which can be mitigated with prior warning of the upstream solar wind conditions.
However, it is currently extremely challenging to accurately predict the arrival time and internal structure of a CME from coronagraph images alone.
In this study, we take advantage of a rare opportunity to use Solar Orbiter, at 0.5\,au upstream of Earth, as an upstream solar wind monitor.
In combination with low-latency images from STEREO-A, we successfully predicted the arrival time of two CME events before they reached Earth.
Measurements at Solar Orbiter were used to constrain an ensemble of simulation runs from the ELEvoHI model, reducing the uncertainty in arrival time from 10.4\,hours to 2.5\,hours in the first case study.
There was also an excellent agreement in the $B_z$ profile between Solar Orbiter and Wind spacecraft for the second case study, despite being separated by 0.5\,au and 10$^{\circ}$ longitude.
The opportunity to use Solar Orbiter as an upstream solar wind monitor will repeat once a year, which should further help assess the efficacy upstream in-situ measurements in real time space weather forecasting.
\end{abstract}

\section*{Plain Language Summary}

Coronal mass ejections (CMEs) are large eruptions of plasma from the Sun that can significantly disrupt human technology when directed towards Earth. 
Much like weather on Earth, the consequences of these `space weather' events can be lessened with warning of their arrival, e.g., putting satellites into safe mode.
This is usually done by identifying a CME in telescope images, and then predicting if and when it will arrive at Earth.
However, the current forecasting models have large uncertainties in arrival time, and struggle to predict the in situ properties of the CME, which can significantly alter the severity of the event.
In this paper, by taking advantage of a period in March 2022, we were able to use real time measurements from halfway between the Sun and the Earth taken by the Solar Orbiter spacecraft.
This allowed us to predict the arrival time of a CME more than a day before it arrived at Earth, a significant improvement on the current capabilities.
We also show that the Solar Orbiter measurements can be used to constrain a CME propagation model, significantly improving the accuracy and precision of the forecasted arrival time.

\section{Introduction}
Space weather can create significant disruption to human technology both in space and on Earth, including loss of satellites, damage to power grids and communication blackouts \cite{Hapgood2012,Eastwood2017a}.
Fortunately, many of these effects can be mitigated with prior warning, meaning that timely and accurate predictions of the arrival and severity of space weather events are extremely important \cite{Eastwood2017}.

The majority of severe geomagnetic storms at Earth are driven by coronal mass ejections \cite<CMEs,>[]{Richardson2001}, large and complex structures impulsively released from the Sun's corona.
Remote sensing observations of the corona can reveal the release of a CME from the Sun, whose propagation through the ambient solar wind can then be simulated to predict an arrival time at Earth.
While there are many sophisticated solar wind and CME models, they often have arrival time errors of $\pm 10$\,hours, partly due to uncertain estimates of the CME's initial parameters \cite{Riley2018, Wold2018}.
In addition, the interaction between the CME and the ambient solar wind can deflect, deform, erode and rotate the CME, which can also significantly affect the arrival time \cite{Wang2002, Wang2004, Gui2011, Good2019,Desai2020,Stamkos2023}.

Information regarding the orientation of the CME's magnetic field, a primary indicator of geo-effectiveness, is often as valuable as the predictions of arrival time \cite{Owens2020e}.
Knowing the potential impact of the CME, rather than just its arrival time, can limit the number of false positives and make predictions more useful for those commercial applications where there is a high cost of mitigation, e.g., putting a spacecraft into a safe mode.
Currently, information about the internal magnetic structure and plasma parameters of an Earth-directed CME are provided by a number of spacecraft at the L1 Lagrange point between the Earth and Sun: Wind \cite{Ogilvie1997}, Advanced Composition Explorer \cite<ACE,>[]{Chiu1998} and Deep Space Climate Observatory \cite<DSCOVR,>[]{Burt2012}.
However, their placement just ahead of Earth only provides a lead time of around an hour, which is often not long enough to take appropriate mitigating measures.

To address many of the problems with the current prediction framework, future upstream solar wind monitors have been proposed, which would be positioned along the Sun-Earth line further from Earth than L1.
However, such a mission would rely on currently inaccessible solar sail technology or a large constellation of orbiting probes \cite{Heiligers2014}, meaning that there are still open questions regarding the efficacy of these future proposals.
In the case of the probe constellation, it would be useful to know the minimum separation in longitude that would provide continuous prediction capabilities.
In addition, the optimal position for a spacecraft would be a trade-off between improved lead time and the accuracy of the prediction.
For example, placing a spacecraft within Mercury's orbit would give several days lead time, but the solar wind structures seen by the spacecraft may have evolved significantly after travelling to Earth \cite{Rodriguez-Garcia2022,Palmerio2022}.
While several case studies have already shown how an upstream spacecraft can be useful in predicting the arrival time and geo-effectiveness \cite{Lindsay1999,Rollett2014,Kubicka2016,Amerstorfer2018,Pal2023}, such a concept has never been attempted in real time.

In this paper, we take advantage of an unprecedented opportunity to use Solar Orbiter \cite{Muller2020} as a real time upstream solar wind monitor.
During a period between February and March 2022, Solar Orbiter crossed the Sun-Earth line at a heliocentric distance of $0.5\,$au, observing two CME events in situ and providing predictions of the arrival time before they arrived at Earth.
In Section \ref{method}, we outline the operational constraints of the Solar Orbiter mission for this purpose and detail how remote sensing images are used in conjunction with numerical models to predict the arrival time at Earth.
We then present the results of our two CME case studies in Section \ref{arrival}, showing how these measurements can be used to improve the uncertainty in predicted arrival time prior to reaching Earth.
Finally, the similarity of the magnetic field structure at 0.5\,au and Earth is investigated in Section \ref{magfield}, opening up the possibility to predict the sub-structure of a CME event, not just the arrival time.

\section{Methodology}\label{method}
\subsection{Operations}

\begin{figure}
    \begin{center}
    \includegraphics[width=0.7\columnwidth]{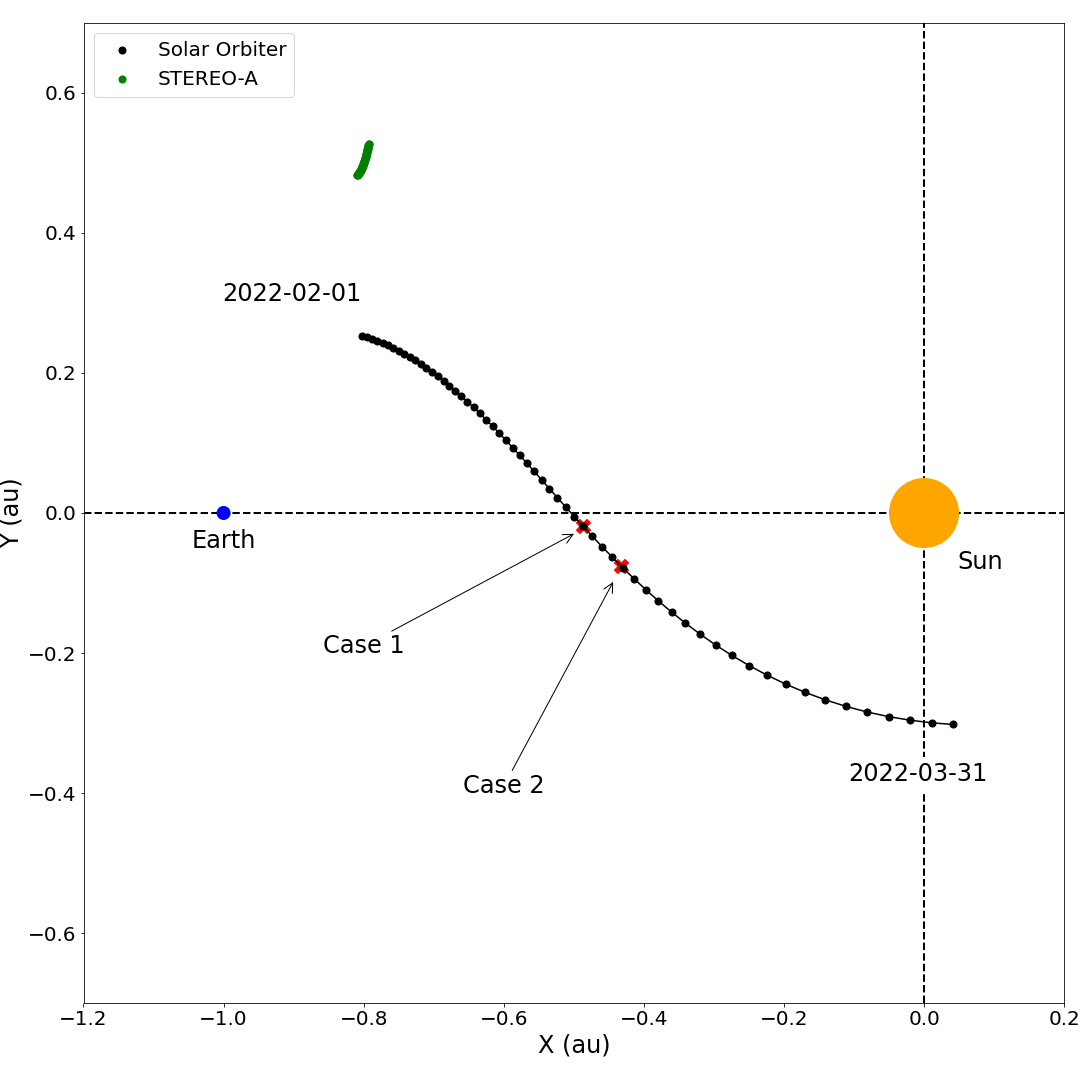}
    \caption{Trajectory of Solar Orbiter in the ecliptic plane of the Sun centred GSE frame between 1 February and 31 March 2022. The spacecraft crossed the Sun-Earth line on 6 March and then subsequently measured the two CME events on 7 March and 11 March, respectively. The relevant timings of each CME event are shown in Table \ref{table}. Each point represents the Solar Orbiter position at the start of each day. The Sun is represented by the orange filled circle, the Earth as a blue dot.}
    \label{fig:traj}
    \end{center}
\end{figure}

During February and March 2022 Solar Orbiter travelled from $0.8\,$au to $0.32\,$au heliocentric distance, crossing the Sun-Earth line on 6 March 2022, as summarised by Fig. \ref{fig:traj}.
While the spacecraft has crossed the Sun-Earth line before, for this crossing, Solar Orbiter had the capability to return data sufficiently quickly to predict the onset and severity of geomagnetic storms at Earth.
This was made possible by the low latency data products returned by the instruments at 100~bits/second, which are intended to help with the `very short term planning' of the remote sensing instruments -- pointing them in the direction of relevant solar structures, such as active regions or the polar coronal holes \cite{Muller2020,Auchere2020}.

Since Solar Orbiter was not designed to be an upstream solar wind monitor, there were a few operational constraints that affected our prediction capability.
First, the data was only downloaded in an $8$-hour window per day, referred to as a `pass'.
Therefore, if a CME arrived between passes, we could not provide predictions until the next pass began, which could be up to $16$\,hours later.
Under normal operation, the low latency data taken between passes is downloaded within 30 minutes of the start of the next pass, while within the pass the latency is less than 5 minutes. 
Since this is only intended to be quick look data for short term planning, this data has a lower time resolution and typically still has artefacts that make it unsuitable for science.

To overcome these challenges, the Solar Orbiter MAG team \cite{Horbury2020a} created a pipeline to produce real time data that was closer to science quality.
Housekeeping data, provided in the low latency packets, was used to remove over $50$ different heater signals from the spacecraft, as well as interference from other instruments aboard Solar Orbiter \cite{Angelini2022}.
This custom pipeline was then run on demand during the pass, and could achieve a latency of just $12$ minutes from taking the measurement in the solar wind to being science quality on the ground, which included a $4$-minute light travel time.
While this pipeline enables real time magnetic field data, it is important to note that this does not include a detailed calibration that is usually provided for science quality data.

For the purposes of this study, the real time data was restricted to the magnetic field measurements.
This is due to the fact that the plasma data from the Proton Alpha Sensor \cite<PAS, >[]{Owen2020} was only available at the end of the $8$-hour pass.
For upcoming radial alignments, both plasma and magnetic field data can be made available in real time.

With Solar Orbiter's position at 0.5\,au, our lead time was around $35\,$hours for a $600\,\textrm{kms}^{-1}$ CME at $0.5\,$au.
To compare with solar wind measurements near Earth, we have used data from the Wind spacecraft at L1.

\begin{table}
    \centering    
    \def\arraystretch{1.5} \begin{tabular}{p{5cm}|p{4cm}|p{4cm}}  
        \textbf{Timings} & \textbf{Case 1} &  \textbf{Case 2} \\ 
        \hline
        CME launch time (STEREO-A) & 2022-03-05 18:00 & 2022-03-10 19:30 \\ \hline
        Solar Orbiter in situ observation & 2022-03-07 22:49 & 2022-03-11 19:52 \\ 
        \textit{~~ Distance to Sun} & ~~$0.48\,$au & ~~$0.44\,$au \\ 
        \textit{~~ Angle to Sun-Earth line} & ~~$2.3^{\circ}$ & ~~$9.6^{\circ}$ \\
        \hline
        Earth estimated arrival time (PAS speed) & 2022-03-10 19:54$\pm 6.8\,$hrs & 2022-03-13 10:18$\pm 3.2\,$hrs \\ \hline
        Earth modelled arrival time (ELEvoHI)  & 2022-03-11 01:04 & 2022-03-13 13:01 \\
        ME, MAE &  $+8.1\,$hrs, $10.4\,$hrs & $+2.2\,$hrs, $2.7\,$hrs \\ \hline
Earth modelled arrival time (constrained ELEvoHI)  & 2022-03-10 14:07 & 2022-03-13 10:25 \\
ME, MAE & $-2.4\,$hrs, $2.5\,$hrs & $-0.1\,$hrs, $1.1\,$hrs \\ \hline
Earth arrival time (corrected from Wind to Earth) & 2022-03-10 16:59 & 2022-03-13 10:53 \\
\end{tabular}\caption{Summary of timings (UTC) and spacecraft position for the two CME case studies investigated. As well as constraining the ELEvoHI model, we also used the CME speed from Solar Orbiter PAS $\pm 50\,\textrm{kms}^{-1}$ to make a simple prediction. The uncertainty of the ELEvoHI model is described as the mean error (ME) and mean absolute error (MAE) \cite{Verbeke2019}.}\label{table}
\end{table}

\subsection{Modelling}\label{modelling}

After identifying a CME within the Solar Orbiter in situ data, we then attempted to predict its arrival time at Earth. 
As well as providing an estimate simply using distance divided by speed from PAS, which inherently neglects effects such as drag, we also applied the ELEvoHI model in real time \cite{Rollett2016,Amerstorfer2018}.
With this particular model, it was not just a fortunate line-up between Solar Orbiter and Earth, but the position of STEREO-A was such that it could provide a side view of any CMEs directed towards Solar Orbiter.
A full description of this model, and the underlying assumptions, can be found in \citeA{Amerstorfer2021} and \citeA{Bauer2021}.

\begin{figure}
    \includegraphics[width=\columnwidth]{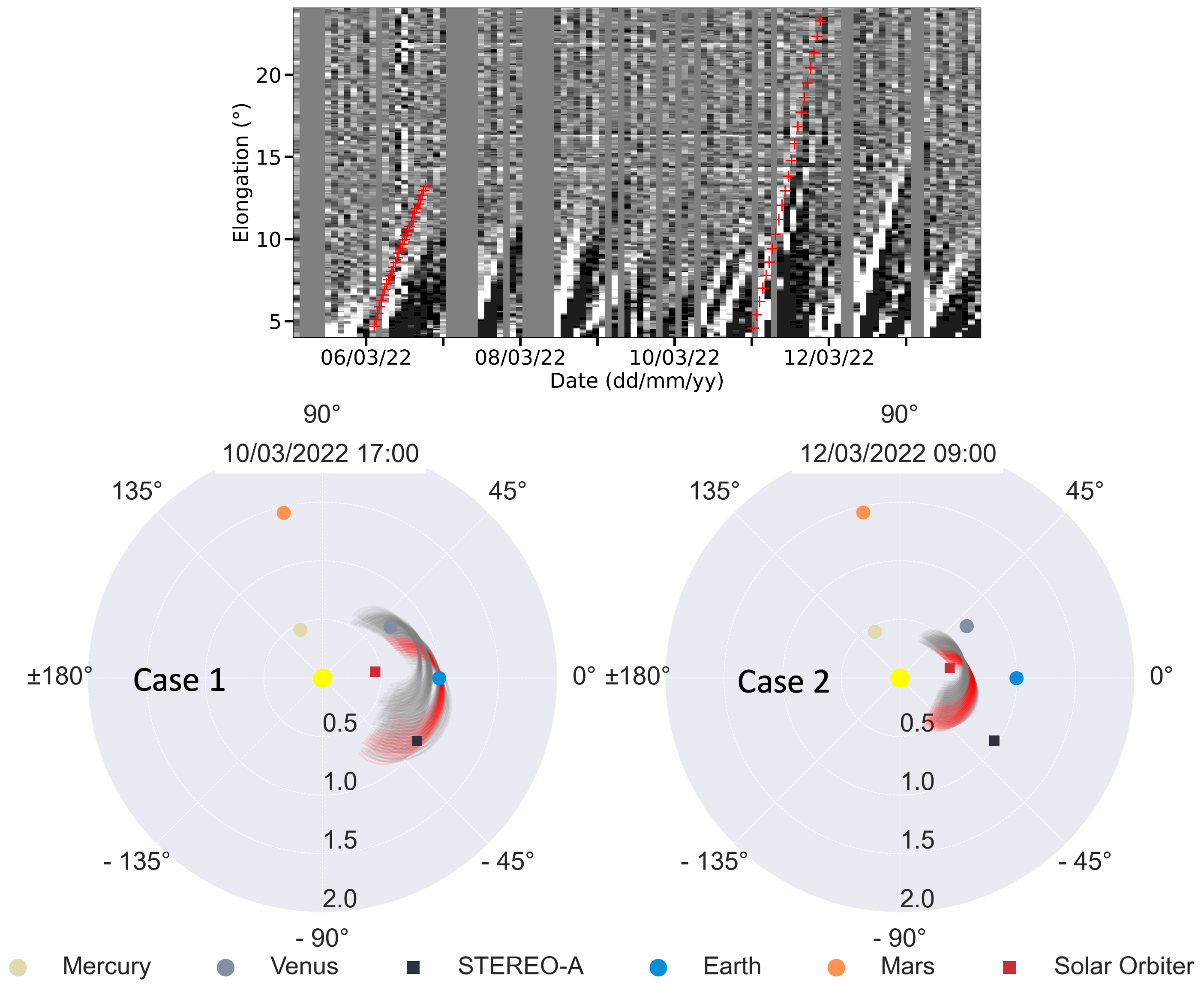}
    \caption{Top panel shows the CME tracks (red) in the STEREO-A HI beacon data, which was used to drive the ELEvoHI model for case 1 (left lower panel) and case 2 (right lower panel). The CME fronts of the 210 ensemble members are shown in the HEE coordinate system. Those ensemble members that did not reach Solar Orbiter within $\pm 4\,$hours of the observed arrival time were rejected (grey), with the remaining runs shown in red.}
    \label{fig:ensemble}
\end{figure}

In essence, this model first uses the heliospheric imager \cite<HI,>[]{Eyles2009} data from STEREO-A to obtain an elongation track that is then converted to radial distance using the ELlipse Conversion method \cite<ELCon,>[]{Rollett2016}.
Tracing the CME front, and fitting the time-distance profile with a drag based model \cite<DBM,>[]{Vrsnak2013}, allows for the estimation of the CME kinematics, which can then extrapolated to Earth with the ELEvo model \cite[]{Mostl2015}.
This framework also requires the CME direction, which is provided by the Fixed-$\phi$ fitting (FPF) model \cite{Sheeley1999}, giving a total of five inputs to the ELEvoHI model.
An ensemble of 210 model runs are created by varying: the direction of CME motion, CME aspect ratio, half-width (according to \citeA{Bauer2021}).
The half width (within the ecliptic) is varied between 25$^{\circ}$ and 45$^{\circ}$ in (steps of 5$^{\circ}$), the direction of motion is varied between 14$^{\circ}$ and 54$^{\circ}$ (relative to STEREO-A) with steps of 2$^{\circ}$.
This roughly corresponds to a direction of $\pm$20$^{\circ}$ longitude.
The inverse ellipse aspect ratio (corresponding to the curvature of the front) is taken as either 0.7 or 0.9 \citeA{Bauer2021}.
These ensemble runs, which can be seen in Fig. \ref{fig:ensemble}, provide a range of arrival times that are treated as the uncertainty in the overall ELEvoHI ensemble model.
For real time applications the beacon, rather than science, data from STEREO-A must be used.
Although this is not as high quality, \citeA{Bauer2021} demonstrated that it can still provide an arrival time prediction with a mean absolute error (MAE) of $11.4\pm8.7$\,hours, as opposed to $8.8\pm3.2$\,hours using science data.

In theory, this prediction uncertainty can be significantly reduced by rejecting those ensemble members that do not match the arrival time of the in situ measurements (with some tolerance) from a spacecraft within $1\,$au.
Constraining a CME model with in situ data has been successful in previous studies, although, these were carried out in hindsight of the CME event \cite{Rollett2014, Kubicka2016, Amerstorfer2018}.
However, with the real time availability of Solar Orbiter data we have, for the first time, constrained the ELEvoHI model with measurements at $\sim$0.5\,au before the CME arrived at Earth.

\section{Results and Discussion}\label{results}

As Solar Orbiter crossed the Sun-Earth line, two CMEs were observed by the spacecraft, at the times depicted as red dots in Figure \ref{fig:traj}.
Both case studies were then subsequently observed by Wind at L1, with the specific timings listed in Table \ref{table}.
We will first discuss the accuracy of the arrival time predictions in Section \ref{arrival}, before investigating the similarity in the magnetic structure between the Solar Orbiter and Wind in Section \ref{magfield}.

\subsection{Arrival Time}\label{arrival}

After observing the first CME event in situ with Solar Orbiter at 2022-03-07 22:49 (case 1), we then tracked the CME front in the time-elongation maps generated from STEREO-A HI beacon data, from 6 March onwards.
Following the method of \citeA{Bauer2021}, this procedure was repeated four more times and interpolated to an equally spaced time axis.
This produced a single profile, as seen in the upper panel of Fig. \ref{fig:ensemble}, that was input into the ELEvoHI model to simulate the CME propagation towards Earth.

\begin{figure}[ht!]
\begin{center}
    \includegraphics[width=0.7\columnwidth]{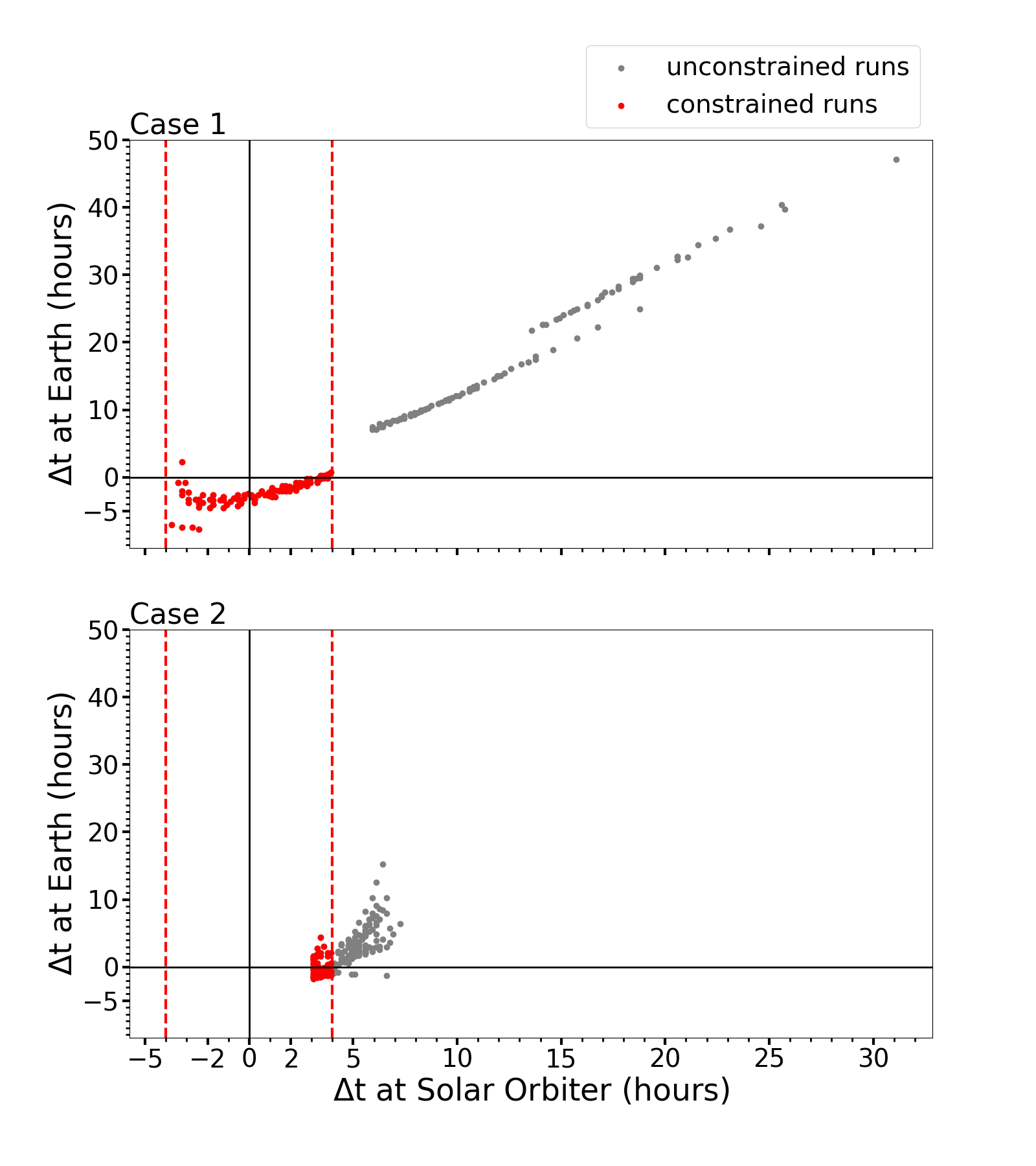}
    \caption{Difference between simulated and true arrival time at Solar Orbiter (x axis) and Earth (y axis) for the 210 ensemble member runs from ELEvoHI. By constraining the model with the Solar Orbiter arrival time $\pm$4\,hours (red dashed lines), only 99 and 85 ensemble members were kept for the two cases, respectively (red scatter points). Therefore, this method improved the accuracy and precision of the final arrival time prediction at Earth.}
    \label{fig:deltaT}
\end{center}
\end{figure}

As discussed in Section \ref{modelling}, an ensemble of 210 model runs are used to make the final prediction of arrival time, as shown in lower panel of Fig. \ref{fig:ensemble}.
The difference between the simulated and true arrival times over the ensemble are shown in Fig. \ref{fig:deltaT}, ranging from -7.8\,to\,47.1\,hours (negative values indicate the prediction was earlier than the true arrival time).
The mean error (ME) in arrival time at Earth was +8.1\,hours, with a mean absolute error (MAE) of 10.4\,hours (Table \ref{table}), representing typical values for such a simulation \cite{Bauer2021,Amerstorfer2021}.

Using the arrival time at Solar Orbiter as a constraint, with a $\pm 4\,$hour threshold, we were able to reject 111 of 210 ensemble members, leaving only 99 runs (red) in Figs. \ref{fig:ensemble} and \ref{fig:deltaT}.
For case 1, this significantly improved the accuracy and precision of the simulation, lowering the ME to -2.4\,hours and the MAE to 2.5\,hours.
In addition, by removing many of the erroneous runs, the range of arrival times was now between -7.7\,to\,2.3\,hours.
While constraining simulations in this way has been attempted in previous studies \cite{Kubicka2016, Amerstorfer2018}, we have shown that this can lead to a drastically improved prediction in real time.

This same method was also applied to case 2, where the ME was reduced from +2.2\,to\,0.1\,hours and the MAE improved from 2.7\,to\,1.1\,hours.
Although this was not done in real time, but still based on HI beacon data, it provided another successful demonstration of the benefits of constraining the ensemble runs.
Such an improvement in arrival time was achieved even when the Solar Orbiter spacecraft was $9.6^{\circ}$ away from the Sun-Earth line.
While this is only one example, it does suggest that a constellation of orbiting probes can be separated by at least $\sim$20$^{\circ}$ at 0.5\,au and still provide useful prediction capabilities.

Even without a complex model, but only using an estimated speed from PAS, we were still able to produce accurate arrival time predictions (Table \ref{table}).
This is likely due to the fact that CMEs are relatively unaffected between 0.5\,to\,$1\,$au, with any significant deflections having already mainly occurred closer to the Sun \cite{Savani2010,Gui2011}.
Of course, CMEs are still known to deflect and deform in the solar wind depending on the downstream conditions \cite{Owens2020c,Desai2020,Hinterreiter2021a,Davies2021a}.
Such a problem could be addressed in future with an improved ELEvoHI model \cite{Hinterreiter2021}, or with a 1D model that can capture CME deformation \cite{Owens2020d}.

\subsection{Magnetic field structure}\label{magfield}

Knowledge of the upstream CME conditions, namely proton speed, density and $B_z$, is arguably of equal importance as the arrival time \cite{Owens2020e}.
Such parameters influence the dynamic pressure of the CME and its ability to trigger magnetic reconnection at the magnetopause \cite{Vasyliunas1982}, which leads to the onset of geomagnetic storms.
The orientation of the magnetic field in the CME, whether $B_z$ is positive or negative, is the primary indicator of storm severity \cite{Vasyliunas1982,Tsurutani1988,Gonzalez1999}, although predicting this property remains a challenging problem for CME models \cite{Savani2015,Kay2017,Mostl2018,Sarkar2020,Reiss2021,Pal2022}.
Therefore, it would be a major advantage if the transient structures seen by a spacecraft within $1\,$au were correlated with those subsequently seen at Earth.

\begin{figure}
    \begin{center}
    \includegraphics[width=0.8\columnwidth]{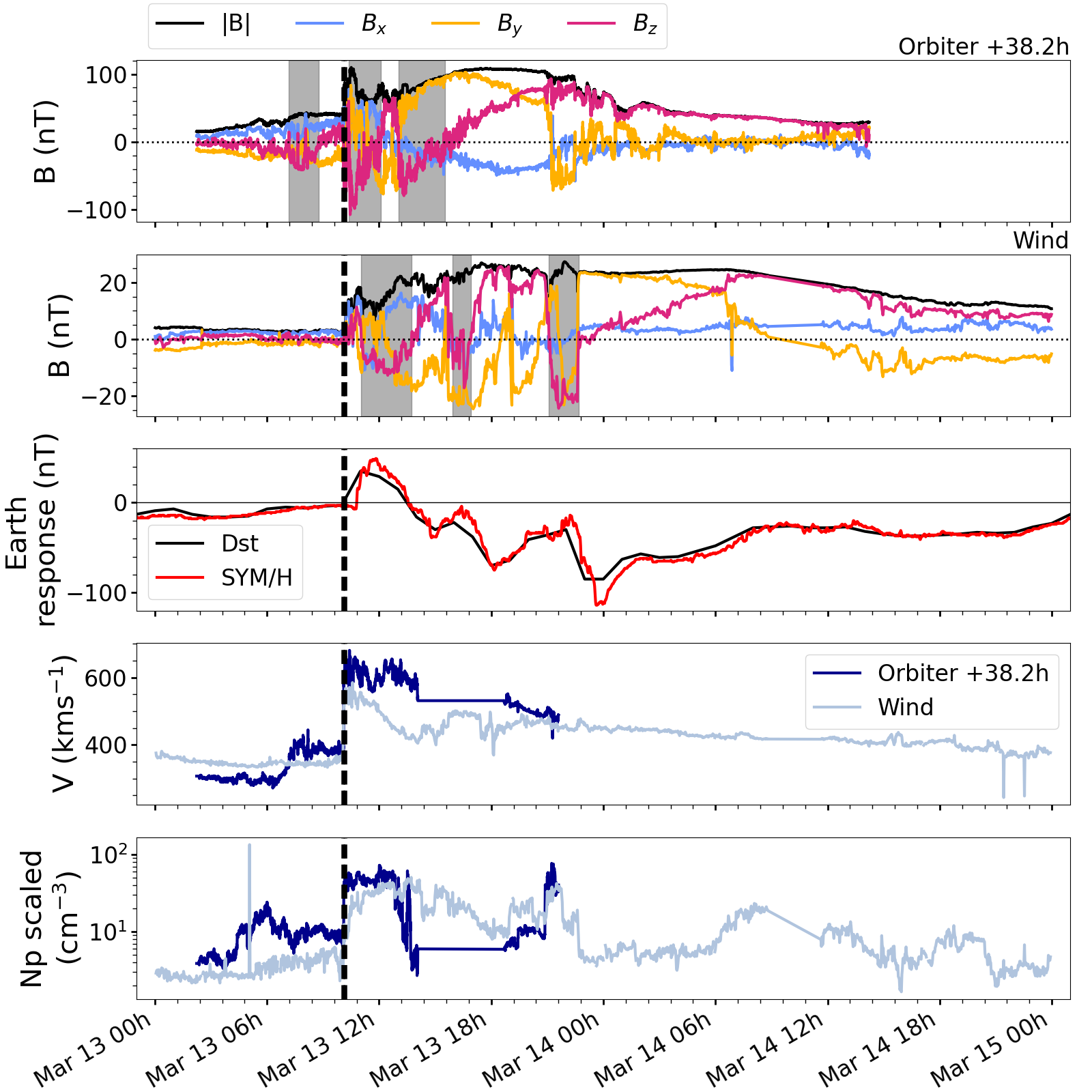}
    \caption{In situ measurements for case 2 in the GSE coordinate system, where vertical dashed lines show the start times of the event. Top panel shows the magnetic field components at Solar Orbiter, which have been shifted by 38.2\,hours to match with the shock front at Wind in the panel below. The proton density, $Np$, at Solar Orbiter has been scaled by $1/r^{2}$. Both spacecraft observe three regions of negative $B_z$, followed by a flux rope with a positive $B_z$. This led to a geomagnetic storm at Earth, which also exhibited three dips in the D$_{ST}$ and SYM/H indices. We conclude that, for this case, the prediction from Solar Orbiter could resolve the small scale changes in the Earth response on the order of hours.}
    \label{fig:big}
    \end{center}
\end{figure}

To evaluate the similarity in structure between the spacecraft, Fig. \ref{fig:big} shows the in situ measurements for case 2 from both Solar Orbiter and Wind, with the former time shifted to match up the shock fronts.
Both spacecraft depict a typical CME profile, with a shock front occurring before a denser and more variable sheath region that was followed by the smooth rotation of the magnetic field in the flux rope.
There was an excellent agreement in the $B_z$ profile between the two spacecraft, with three periods of negative $B_z$ (highlighted regions) followed by a flux rope with a mainly northward orientation.
The first $B_z<0$ region in Solar Orbiter is now part of the CME sheath at Wind, having been overtaken by the shock.
This implies that this structure is coherent over 10$^{\circ}$ longitude between Solar Orbiter and Wind, which along with its increased speed over the background, suggest that it could be a stream interaction region, or perhaps a small CME.

The overall $B_z$ profile was relatively unchanged at Wind due to the low inclination of the shock at Solar Orbiter, which had an azimuth ($\phi$) of $56^{\circ}$ and an elevation ($\theta$) of $4.4^{\circ}$ found using the cross product of the magnetic field either side of the shock. 
Such angles are in the RTN coordinate system, where $\phi$ is the angle in the R-T plane where $0^{\circ}$ points along $\vec{R}$ and $90^{\circ}$ along $\vec{T}$. $\theta$ is the angle out of the R-T plane.
It is important to note that the shock can only scale the different magnetic field components, so is not responsible for any changes in the sign of the components between the two spacecraft.

\begin{figure}[hb!]
    \begin{center}
    \includegraphics[width=0.7\columnwidth]{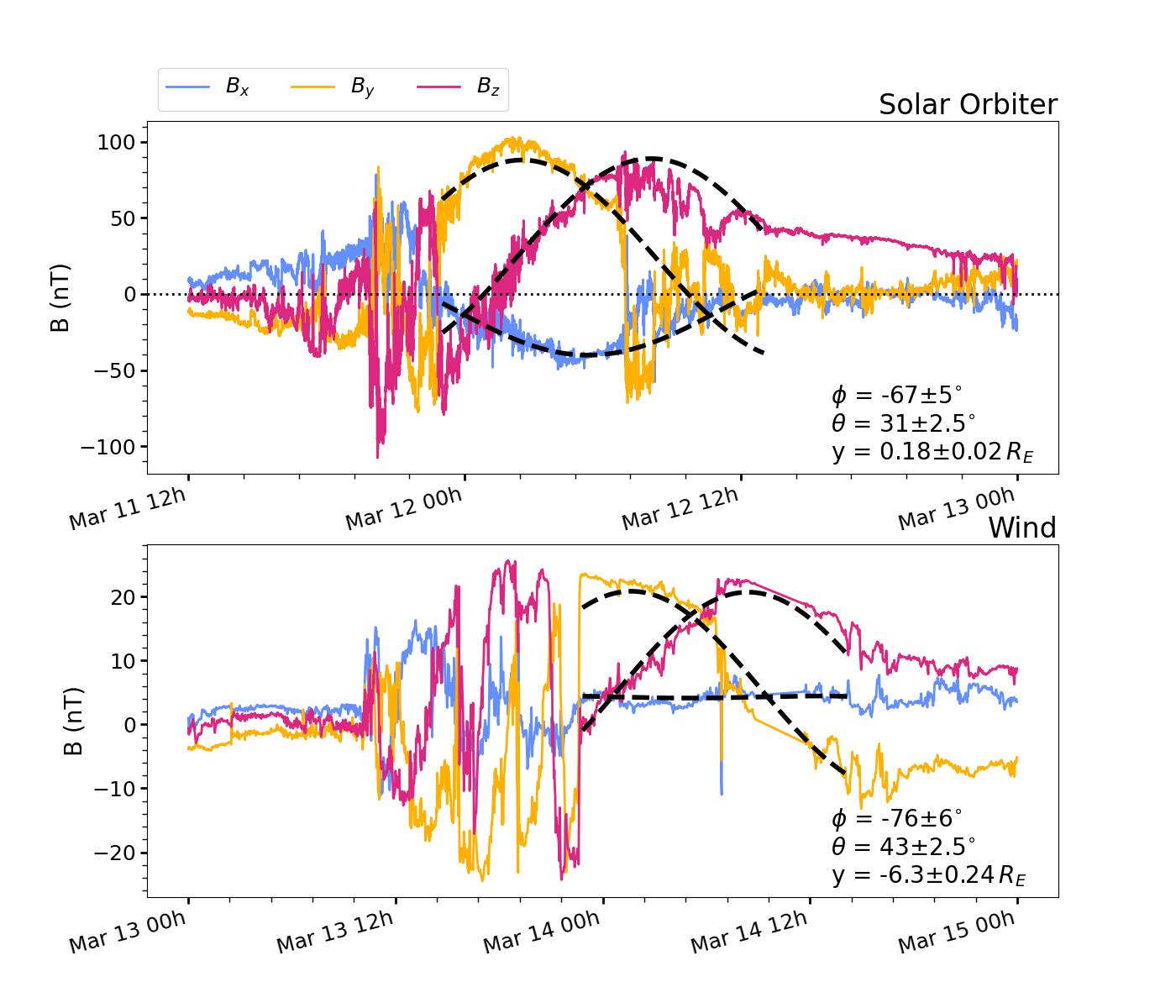}
    \caption{Flux rope fits for case 2 at Solar Orbiter (top) and Wind (bottom), with the elevation ($\theta$), azimuth ($\phi$) and impact factor (y) shown in the lower right. These fits demonstrate that the flux rope had a consistent orientation between the spacecraft, with a differing impact factor creating an altered profile for $B_y$ and $B_z$.} 
    \label{fig:flux_rope_fit}
    \end{center}
\end{figure}

The $B_z$ signature was also comparable in the CME flux rope, although, this was less clear for the other magnetic field components shown in Fig. \ref{fig:flux_rope_fit}.
To further assess the similarity between the two CMEs, we fit the flux rope signatures with a simple force-free model assuming cylindrical symmetry and using Bessel function solutions \cite{Burlaga1988,Lepping1990}. 
We find a helicity sign of -1 for both events, and an azimuth, elevation and impact factor of ($-67\pm5^{\circ}$, $31\pm2.5^{\circ}$, $0.18\pm0.02~R_E$ (0.8\%)) for Solar Orbiter and ($-76\pm 6^{\circ}$, $43\pm2.5^{\circ}$, $-6.3\pm0.24~R_E$ (19\%)) for Wind.
While the fitting is sensitive to variations in CME boundary definition, the results are consistent enough to demonstrate that the orientation of the flux rope remains fairly stable, with the changes in magnetic field component being mostly due to a change in impact factor.
Such a scenario is consistent with the relative positions of the two spacecraft, with Solar Orbiter being $9.6^{\circ}$ away from the Sun-Earth line at 0.5\,au and $\sim 3^{\circ}$ higher in latitude.
Given that $B_x$ and $B_z$ were both consistent in the sheath region the change of sign of impact factor may also explain the differences in the $B_y$ component during this time, with Wind showing a predominantly negative $B_y$ compared to Solar Orbiter.

In summary, both the orientation of the CME sheath and flux rope were similar between the spacecraft for case 2, demonstrating that CME structures can remain coherent between an upstream monitor and Earth.
Interestingly, the magnetic response at Earth (D$_{ST}$ and SYM/H indices) was consistent with the shape of the $B_z$ signature at Solar Orbiter, displaying three dips before the flux rope slowly rotated northward.
So, at least for this event, the magnetic storm indicators at Earth were strongly correlated to the magnetic structure seen at 0.5\,au around 40\,hours prior.
This suggests that with an upstream spacecraft, it could be possible to not only predict the arrival of a CME, but the sub-structure of the flux rope and sheath region using measurements from 0.5\,au.

For case 2, the magnetic field at Solar Orbiter was sufficient to capture the general trends in D$_{ST}$ at Earth, although, there were still some clear differences in Fig. \ref{fig:big}, e.g. differing field magnitude strength between the shaded regions.
Much like the improvements in arrival time from Section \ref{arrival}, knowledge of the in situ CME properties, such as shock or flux rope orientation, could be used as the initial parameters for a CME propagation model \cite{Sarkar2020}.
Again, this would not have to account for deflections in the corona, and should make for more accurate predictions of magnetic structure compared to those based on coronagraph observations.
This would be another step towards being able to predict how the geomagnetic storm at Earth will develop on an hourly timescale.

In addition, an upstream solar wind monitor would be especially useful for simulating the pileup of material in the sheath region \cite{Kay2022}, which is known to be major drivers of geomagnetic storms \cite{Tsurutani1988,Gonzalez1999,Lugaz2016}.
Not only could the shock orientation be measured in situ, enabling the application of the Rankine-Hugoniot conditions, but the ambient solar wind could be continuously monitored.

\begin{figure}
    \begin{center}
    \includegraphics[width=0.8\columnwidth]{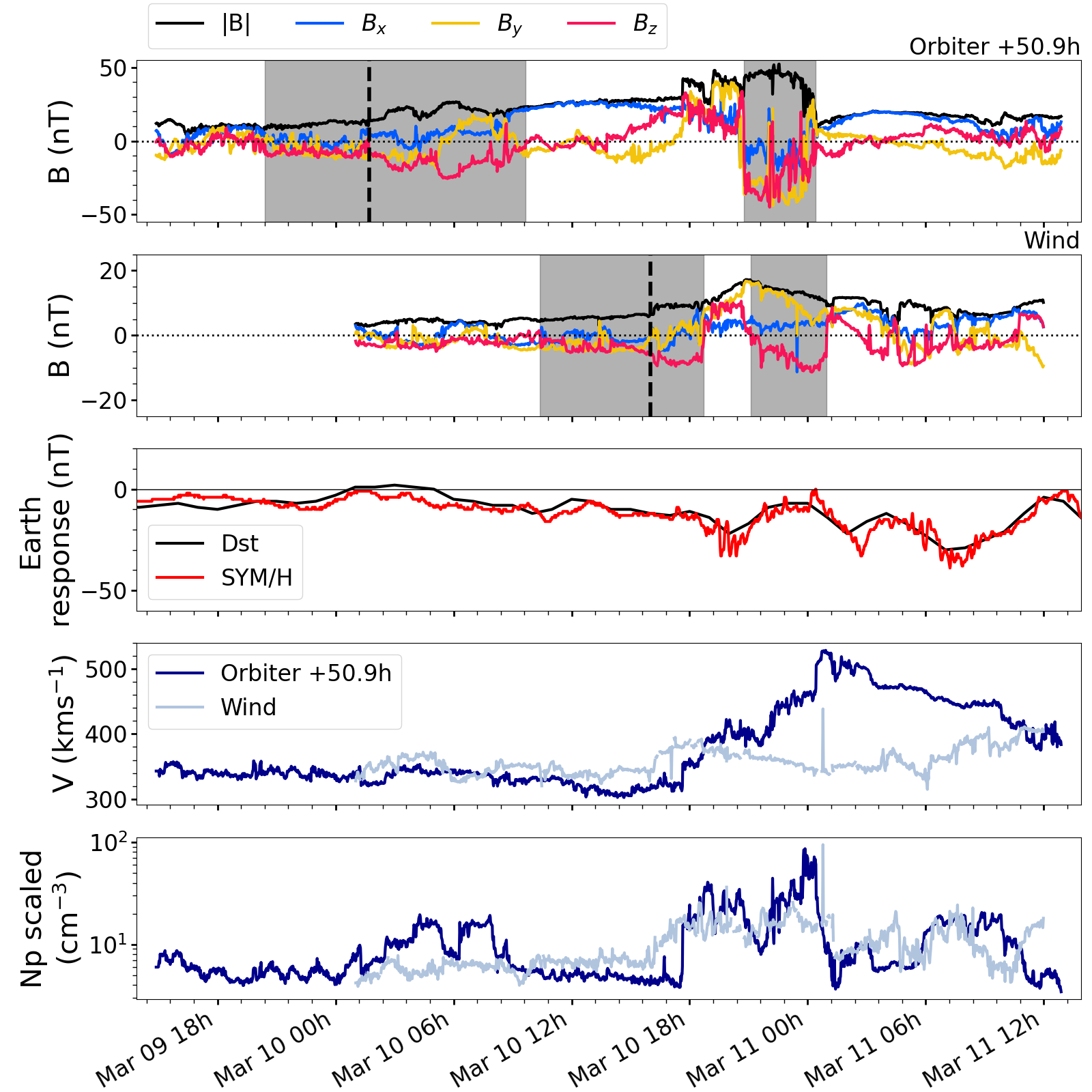}
    \caption{In situ measurements of the double flux rope event in case 1, where vertical dashed lines show the start times of each event. Top panel shows the magnetic field components for Solar Orbiter, which have been shifted by 50.9\,hours to match the Wind observations near Earth in the panel below. Grey highlighted regions show periods of negative $B_z$, which had potential to generate a geomagnetic storm. However, the D$_{ST}$ and SYM/H indices (middle panel) showed that this was only a minor storm.}
    \label{fig:double}
    \end{center}
\end{figure}

It is important to note that this is only a single CME event, and a more extensive statistical study is needed.
Indeed, the in situ structure of case 1, in Fig. \ref{fig:double}, was a complex interaction of two flux ropes, rather than the typical sheath and flux rope profiles seen in case 2.
It is clear that there has been some interaction between the two flux ropes by the time event appears at Wind.
However, both flux ropes still appear to hold their own structure, which combined with the large separation in time, suggests that the interaction between them was not strong.
In addition, the ELEvoHI model only models the evolution of the density enhancement visible in HI images, rather than modelling the magnetic structure within the ejecta.
While a detailed post-analysis might offer insights into these interactions, it's important to remember that our focus was on real-time prediction: a proof of concept that, by its nature, operates under constraints of limited data and time.

\section{Conclusions}

In this study, we took advantage of an opportunity to use Solar Orbiter as an upstream space weather monitor between February and March 2022.
As Solar Orbiter crossed the Sun-Earth line, we were able to use in situ data on the ground which was taken in the solar wind only 12 minutes earlier. 
In combination with the favourable position of STEREO-A, we were able to model the kinematics of two CME events with the ELEvoHI model. 

We first demonstrated how knowledge of arrival time at Solar Orbiter could be used to constrain the ELEvoHI model.
Under normal operation, this model uses an average of 210 ensemble members to estimate the arrival time of a CME at Earth.
However, by only keeping those ensemble members that were within $\pm4$\,hours of the arrival time at Solar Orbiter, we improved both the accuracy and precision of the model.
Specifically, the MAE was reduced from 10.4\,to\,2.5\,hours and 2.7\,to\,1.1\,hours in the two case studies.
Therefore, for the first time, a numerical model was constrained with data from 0.5\,au to produce an updated prediction before the actual arrival of the CMEs at Earth.
As well as demonstrating how a spacecraft at 0.5\,au could provide a lead time of over 40\,hours, we also showed that the predictions were still accurate when Solar Orbiter was $9.6^{\circ}$ away from the Sun-Earth line.
Such a result provides motivation for a future constellation mission housing in situ instrumentation, since the individual spacecraft could be separated by $\sim 20^{\circ}$, making the concept more viable.
While we could have used another CME model, we found it particularly beneficial to model the CMEs with the aid of STEREO-A HI data away the Sun-Earth line.
Similar dedicated real time HI data can hopefully be provided by ESA's Vigil mission in the near future.

We also found that even simple estimates of arrival time, using just the average CME speed at Solar Orbiter, could produce accurate arrival times at Earth.
This represents a major benefit of an upstream solar wind monitor, as it reduces the need to model complex interactions and deflections in the corona, that can drastically alter arrival time.
Of course, there is still a need to account for compression, distortion, erosion and rotation as the CME propagates in the solar wind.
This, and the interaction between CMEs, could be handled by a 1D model \cite{Owens2020d} or an improved version of ELEvoHI in the future \cite{Hinterreiter2021}.

Comparing measurements from Solar Orbiter and Wind revealed that the magnetic structure of the CME sheath and flux rope was remarkably similar for the second case study, despite being separated by 0.5\,au.
Crucially, these periods of negative $B_z$ were also seen to match well with the magnetic response at Earth (D$_{ST}$ and SYM/H profiles).
While more modelling work is needed in the future, this is an encouraging indicator for the possibility to predict the evolution of a geomagnetic storm using measurements at 0.5\,au.
This would also be important for reducing the number of false positive predictions, since the geo-effectiveness could be evaluated more than a day in advance.

In future, we hope that more models can make use of this data, either to constrain the output or to initiate a CME simulation away from the complex environment near the Sun.
Fortunately, the opportunity to use Solar Orbiter for this purpose repeats once a year, with more CMEs being released as the Sun approaches solar maximum.

\section*{Open Research Section}
The data used in this paper are available at the following places: Solar Orbiter data can be found on the Solar Orbiter archive \cite{SOAR}; Wind data can be found at the Space Physics Data Facility \cite{NASA2023a}; quick look D$_{ST}$ data is available from \cite{WorldDataCenterforGeomagnetism2015}; the SYM/H data is available from \cite{WorldDataCenterforGeomagnetism2022} and STEREO-A HI data are available from \cite{UKSolarSystemDataCentre2023}.

The ELEvoHI model is available on GitHub \cite{Amerstorfer2020}.

\acknowledgments
RL was supported by an Imperial College President’s Scholarship and TSH by STFC ST/S000364/1. 
The Solar Orbiter magnetometer was funded by the UK Space Agency (grant ST/T001062/1). We acknowledge the work of all the engineers who supported the instrument development and calibration, together with the engineering and technical staff at the European Space Agency, including all the Solar Orbiter instrument teams, and Airbus Space. T.A and M.B thank the Austrian Science Fund (FWF): P 36093, P31659. C.M. and E. E. D. are funded by the European Union (ERC, HELIO4CAST, 101042188). The HI instruments on STEREO were developed by a consortium that comprised the Rutherford Appleton Laboratory (UK), the University of Birmingham (UK), Centre Spatial de Li\`ege (CSL, Belgium) and the Naval Research Laboratory (NRL, USA). The STEREO/SECCHI project, of which HI is a part, is an international consortium led by NRL. J.D., R.H and D.B. recognise the support of the UK Space Agency for funding STEREO/HI operations in the UK. M.D. acknowledges the support by the Croatian Science Foundation under the project IP-2020-02-9893 (ICOHOSS). Views and opinions expressed are however those of the author(s) only and do not necessarily reflect those of the European Union or the European Research Council Executive Agency. Neither the European Union nor the granting authority can be held responsible for them.

\bibliography{library}

\end{document}